\documentclass[12pt,aps,prb,preprint]{revtex4}
\usepackage{graphicx}
\usepackage{dcolumn}
\usepackage{amsmath}
\usepackage{amssymb}
\usepackage{float}
\usepackage{epstopdf}

\def\rv{{\bf r}}

\def\beq{\begin{equation}}
\def\eeq{\end{equation}}

\begin{document}

\title{Evaluation of inverse integral transforms for undergraduate physics students}
\author{Aaron Farrell, Brandon P.~van Zyl and Zachary MacDonald} 
\affiliation{Department of Physics, St. Francis Xavier University, 
Antigonish, NS, Canada B2G 2W5} 
\begin{abstract}
We provide a simple approach for the evaluation of inverse integral transforms that does not
require any knowledge of complex analysis.  The central idea behind the method is to
reduce the inverse transform to the solution of an ordinary differential equation.
We illustrate the utility of the approach by providing examples of  the evaluation of  transforms, without the use of tables.
We also demonstrate how the method may be used to obtain a general representation of a function in the form of a series involving the Dirac-delta distribution
and its derivatives, which has applications in quantum mechanics, semi-classical, and nuclear physics.   
\end{abstract}
\maketitle
\section{Introduction}
Integral transforms are usually presented to undergraduate physics majors well before they are exposed to them in any rigorous fashion in their mathematics courses. The primary motivation behind the integral transform is to re-write an ODE given in one representation (the original $t$-domain), to another representation (the $s$-domain, or image domain) where the solution is hopefully simpler; $t$ and $s$ are called conjugate variables.  Once the solution in the $s$-domain is known, the inverse transform is applied to obtain the desired solution in the original $t$-representation.  In this way, the integral
transform is presented as yet another tool to add to the arsenal of techniques that may be applied to find the solution of linear ordinary differential equations.

The usefulness of the integral transform method for the solution of an ODE (or a set of ODEs) is clearly predicated on the ability to perform the inverse transformation back to the original domain.  However, if the
inverse transform requires complex contour integration methods ({\it e.g.,} the inverse one-sided Laplace, two-sided Laplace,  Mellin, and Weierstrass transforms), most undergraduate physics students
will find themselves at an impasse; namely, students learning about methods for solving linear ODEs will likely have had no exposure to complex analysis, and therefore, be unable to perform the inverse transformation without the use tables, or perhaps  symbolic mathematics packages such as Maple$^\copyright$ or 
Mathematica$^\copyright$.  If the inverse transform cannot be obtained, the student is left with no choice but to attempt to solve the original problem using the set of techniques presented to them 
in their course(s) on linear ordinary differential equations.    

We wish to stress from the outset that the outcome of this tutorial {\em does not} change the situation described above; that is, if the problem
is to find the solution to an ODE via an integral transform, and the inverse transform is not tabulated anywhere, then this article will be of little help.
Rather, we have in mind a scenario in which a particular quantity of interest is explicitly formulated in terms of an inverse integral transform, and where the evaluation of the inverse transform
ostensibly requires the use of complex contour integration methods.  
Our target demographic then, are undergraduate physics students with no exposure to complex analysis, but already familiar with the solution of linear ordinary differential equations.

To this end, the rest of our paper is organized as follows.  In the next section, we provide a brief overview of general linear integral transforms, from which we focus our attention to the most common transforms
arising in physics, where the inverse transform is a complex contour integral.   We then present some simple rules which may be applied to reduce the complex contour integration problem to the
solution of a linear ODE.~\cite{note1}  In Sec.~III, we illustrate the method by providing several examples 
of inverse transforms involving complex contour integrals,  which are either not easily found in standard references, or not readily obtained using symbolic mathematics packages.  
In all of our examples, the emphasis is on {\em how to find} the inverse transform, and not technical mathematical details, 
which we are confident will be covered in more formal mathematics courses.   In Sec.~IV, we demonstrate how our approach
may also be applied  to obtain a multipole expansion of a function, which would otherwise require a much more extensive mathematical analysis. 
Finally, in Sec.~V we close the tutorial with some  concluding remarks.
\section{Integral transforms}
We consider a generic integral transform of the form~\cite{arfken}
 
 \begin{equation}\label{defi}
 f(s) = \mathcal{I}[F(t)] \equiv \int_{t_1}^{t_2}K(s,t)F(t)dt~, 
 \end{equation}
 where $t \in \mathbb{R}$, $K(s,t)$ is the kernel, and $f(s)$ the image of $F(t)$ under the transform $\mathcal{I}$. The inverse transform is given by
 
   \begin{equation}\label{inver}
 F(t) = \mathcal{I}^{-1}[f(s)] \equiv \int_{s_1}^{s_2}K^{-1}(s,t)f(s)~ds~,
 \end{equation} 
 such that 
 
 \begin{equation}\label{ortho}
 \int_{s_1}^{s_2} K^{-1}(s,t)K(s,t')ds= \delta(t-t'). 
 \end{equation}
Equation (\ref{ortho}) may be used to determine the inverse Kernel, $K^{-1}(s,t)$ and, therefore whether the parameter, $s$, is real or complex.
 
We now recall some properties of the integral transform. Linearity, $\mathcal{I}[F(t)+F(t)] = \mathcal{I}[F(t)]+\mathcal{I}[G(t)]$, and multiplication by a constant $\mathcal{I}[AF(t)] = A\mathcal{I}[F(t)]$ follow from the integral definition of the transform given in Equation (\ref{defi}).  Furthermore, it also follows immediately from Eq.~(1) that
 
 \begin{equation}
 sf(s) = \int_{t_1}^{t_2}(sK(s,t))F(t)~dt ;\;\;\;\;\;\; f^{(1)}(s) = \int_{t_1}^{t_2}\frac{\partial K}{\partial s}F(t)~dt,
 \end{equation} 
 where the superscript in parenthesis denotes the order of the derivative with respect to the argument of the function.
 
 For the purposes of this paper, let us now focus on a class of Kernels (see also Sec.~III C below), given by
 \beq
 K(s,t) = Ce^{\alpha st},~~~~~~\alpha,~C \in \mathbb{C}
 \eeq
 which have the properties 
 \beq 
 \frac{\partial K}{\partial s} = \alpha t K~,~~~~sK = \frac{1}{\alpha} \frac{\partial K}{\partial t}~.
 \eeq
 We thus observe that multiplication by $s$ amounts to differentiating with respect to $t$ and derivatives with respect to $s$ amount to multiplying by $t$.
 It is a straightforward exercise to show that after $n$-fold applications of the multiplication and differentiation rules, we obtain

\beq 
s^{n}f(s) = \frac{\bigtriangleup_{n}}{\alpha^n} + \frac{(-1)^n}{\alpha^{n}}\mathcal{I}[F^{(n)}(t)]~,~~~~f^{(n)}(s)= \alpha^{n}\mathcal{I}[t^{n}F(t)]~,
\eeq
where 
\beq
\bigtriangleup_n \equiv \Big[\sum_{m=1}^{n}(-1)^{m}(s\alpha)^{n- m}e^{s\alpha t} F^{(m-1)}(t)\Big] \Big|_{t_1}^{t_2}~,
\eeq
are surface terms resulting from $n$-applications of integration by parts. Kernels of the type in Eq.~(5) also possess the so-called ``shifting" property (let us take $C=1$ for clarity),
 
 \begin{equation}
 f\left(s+\frac{A}{\alpha}\right)=\int_{-\infty}^{\infty}e^{\alpha s t}[e^{A t}F(t)]dt = \mathcal{I}[e^{At}{F(t)}]~,
 \end{equation}
and
\begin{equation}
 e^{-As}f(s)=\int_{-\infty}^{\infty}e^{\alpha s t}e^{-As}F(t)dt = \int_{-\infty}^{\infty}e^{\alpha(t-\frac{A}{\alpha})s}F(t)dt = \mathcal{I}\left[F\left(t+\frac{A}{\alpha}\right)\right]~.
\end{equation}

\mbox{}\\ \\
\begin{table}[h]
\begin{center}
\begin{tabular}{| c | c | c | c | c | c | c | c | } \hline
Transform Type& $K(s,t)$ & $t_1$ & $t_2$ &  $K^{-1}(s,t)$ & $s_1$  & $s_2$ & $\bigtriangleup_n$\\ \hline
One-Sided Laplace & ${e^{-st}}$ & $0$ & $\infty$ & $\frac{e^{st}}{2\pi i}$  & $c-i\infty$ & $c+i\infty$ & $\sum_{m=1}^{n}s^{n- m}F^{(m-1)}(0)$  \\  \hline
Two-Sided Laplace & ${e^{-st}}$ & $-\infty$ & $\infty$ & $\frac{e^{st}}{2\pi i}$  & $c-i\infty$ & $c+i\infty$ & 0  \\  \hline
Fourier & $\frac{e^{-ist}}{\sqrt{2\pi}}$ & $-\infty$ & $\infty$ & $\frac{e^{ist}}{\sqrt{2\pi}}$  & $-\infty$ & $\infty$ & 0   \\  \hline

\end{tabular}
\end{center}
    \caption{Integral transforms constructed after making particular choices for the parameters discussed in the text.}
    \end{table}

\begin{table}[h]
\begin{center}
\begin{tabular}{| l | l | l |} 

  \hline
  Property & Image Function & Inverse Integral Transform with $K(s,t)=Ce^{\alpha s t}$\\ \hline
  I & $af(s)+bf(s)$ & $aF(t)+bG(t)$\\ \hline
  II & $f^{(n)}(s)$ & $ \alpha^{n}t^{n}F(t)$\\ \hline
  III & $s^{n}f(s) $ & $ \bigtriangleup_{n} + \frac{(-1)^n}{\alpha^{n}}F^{(n)}(t)$ \\ \hline
  IV & $ e^{-As}f(s)$ & $F\left(t+\frac{A}{\alpha}\right)$\\ \hline
  V & $f\left(s+\frac{A}{\alpha}\right)$ & $e^{ At}{F(t)}$\\ \hline
  \end{tabular}
  \end{center}
  \caption{A summary of some useful properties of integral transforms with Kernel $K(s,t)=Ce^{\alpha s t}$}
  \end{table}

In Table I, we present the various transforms that may be constructed by making particular choices for the parameters discussed above.  Table II provides a summary of some of the
properties of integral transforms with Kernels of the form $K(s,t) = Ce^{\alpha s t}$.

\subsection{The method}

We are now in a position to present the general method for reducing the evaluation of the inverse integral transform,  to a linear ordinary differential equation.  The method can be summarized as follows:
\mbox{}\\\\
\begin{tabular}{ l p{15cm} }    
   $1)$ &  Find the ODE or difference equation satisfied by the image, $f(s)$, whose inverse transform, $F(t)$, we desire.\\ 
   $2)$ & Term by term invert the equation obtained in step $1)$ using the properties listed in Table II above.\\
   $3)$ & Solve the resulting ODE/difference equation for $F(t)$, thereby obtaining the desired inverse transform.
\end{tabular}
\mbox{}\\\\
It should now be clear why our method is not particularly useful for the study of ODEs.  In the case of integral transforms applied to ODEs, one {\em starts} in the $t$-domain, looking for the solution, $F(t)$.
The integral transform gives an algebraic equation for $f(s)$, which if solved, is followed by an application of the inverse transform to obtain $F(t)$ in the original domain.  However, if we apply
steps 1) and 2) above to $f(s)$, the result will simply be a restatement of the  
original ODE in the $t$-domain, which is of no help, given that it was the ODE in the $t$-domain that motivated the transform to the $s$-domain in the first place.

Instead, our method is designed to reformulate a problem which may not be easily accessible to a student,
{\it e.g.,} finding the inverse transform via contour integration, to one for which the student is already well equipped to handle --- the solution of a linear ODE.
We are not suggesting that the solution of the resulting ODE will be simple (or even possible), but rather, the key point is that the student will be empowered to find the inverse 
transform using methods they are already comfortable, without having to rely solely on tables.
In the next section we present several applications of the method, which will serve to both clarify its utility and applicability.
\section{Applications}

\subsection{Two-sided inverse Laplace Transform} 

From Table I, the two-sided Laplace transform (LT) is given by~\cite{vanderpol}
\beq
f(s) = \mathcal{B}[F(t)] = \int_{-\infty}^{\infty} e^{-st} F(t)~dt~.
\eeq
It is assumed that  the two-sided Laplace integral, Eq.~(11), is convergent for ${\rm Re}~s = c$;  the question of whether or not the integral may be convergent for other values of $s$ may be
answered by powerful theorems in Laplace transforms, which are not the focus of the present tutorial.~\cite{vanderpol}  The inverse transform is given by 
\beq
F(t) = \mathcal{B}^{-1}[f(s)] = \frac{1}{2\pi i}\int_{c - i \infty}^{c+i\infty}e^{s t} f(s)~ds~.
\eeq
The constant, $c$, must be greater than the largest real part of the zeros of the transform function.  Therefore, for the two-sided inverse Laplace transform (ILT), 
the contour may be closed in either the left or right half plane, the choice being dependent on the function, $F(t)$.

The two-sided LT may in fact be used as the basis for all of the integral transforms presented in Table I.  For example, if one puts $s \to i s$ in Eq.~(11), the Fourier transform (FT) (aside from a constant) is immediately
obtained.~\cite{vanderpol}  The one-sided LT is likewise simply a special case of Eq.~(11) provided we replace $F(t) \to F(t) \Theta(t)$, where $\Theta(t)$ is the unit Heavisde function.   Moreover, the two-sided LT
allows for the treatement of functions defined over $t \in \mathbb{R}$, in addition to loosening the restriction imposed by the FT, namely, that $F(t)$ be absolutely integrable over
$t \in \mathbb{R}$.
It is also evident from Table I that the two-sided LT has very simple operational rules; that is, $\bigtriangleup_n =0$, whereas for the one-sided LT, $\bigtriangleup_n \neq 0$ and requires a careful application
to the term-by-term inversion of the ODE satisfied by $f(s)$ discussed above.

\subsubsection{A few known results}
To see the power of our approach in action, let us first demonstrate how to obtain some well known ILTs~\cite{vanderpol}  by making use of the much simpler two-sided operational rules.  
\mbox{}\\\\
{\bf Example 1}.
Consider the following image
\beq 
f(s) = \frac{1}{s^{n+1}},~~~~n \geq 0~.
\eeq
The problem  is to find the ILT of the image function, which by definition, requires the calculus of residues.  However, if we apply our three-step procedure
outlined above, we find that $f(s)$ satisfies,
\beq
s f^{(1)}(s) + (n+1)f(s)=0~.
\eeq
Using the properties listed in Table II,  we see that the ODE for $F(t)$ is given by
\beq
-t F^{(1)}(t) + n F(t) = 0~,
\eeq
which has the solution
$F(t) = C t^n$.  The constant $C$ is determined by demanding that the two-sided LT of $F(t)$ (a {\em real} single variable integral) is identical to the image function $f(s)$.  It is easy to see
that  we must have 
 \beq
 F(t) = \mathcal{B}^{-1}\left[ \frac{1}{s^{n+1}}\right] = \frac{t^n}{\Gamma(n+1)}\Theta(t)~,
 \eeq
which is a well known result from the one-sided ILT tables.  Note that our simple two sided operational rules naturally reproduce the one-sided
 results, and no complex contour integration has been used.
\mbox{}\\\\
{\bf Example 2}. 
 Next, consider the  image function
 \beq
 f(s) = s^n~,~~~~n \geq 0~,
 \eeq
whose ILT we wish to find using our ODE method. 
The image function satisfies the following first-order ODE,
 \beq
 s f^{(1)}(s) - n f(s) = 0~.
 \eeq	
 Applying the same approach as in the previous example, we find that $F(t)$ satisfies
 \beq
 t F^{(1)}(t) = -(n+1) F(t)~.
 \eeq
 Now, it is easy to show that $F(t) = C/t^{n+1}$  ($C$ is an integration constant) is a solution to Equation (19).  However, 
 \beq
 \int_{-\infty}^{\infty} \frac{e^{-st}}{t^{n+1}}~dt~,
 \eeq
 does not converge for any $n \geq 0$.   Therefore, while we have obtained a valid solution to Eq.~(19), it cannot be the two-sided ILT of $f(s)$ since the forward transform, {\it viz.,} Eq.~(20) {\em does not exist}.
If $F(t)$ cannot be given by the solution of a function, what is the other possibility?  As it happens, the {\em only}
 way to obtain a solution to Eq.~(19), and have the forward transform exist, is to extend the class of functions over which the two-sided LT is defined to so-called {\em tempered distributions}.~\cite{temper}
 Tempered distributions are beyond the scope of this tutorial, but fortunately, in the present case, the required distribution is the well-known Dirac-delta distribution.

Using the fundamental properties for the derivative of the Dirac-delta distribution, we observe that
\beq
 \int_{-\infty}^{\infty} e^{-st} \delta^{(n)}(t)~dt =  s^n~.
 \eeq
We therefore conclude that 
\beq
F(t)=\mathcal{B}^{-1}[s^n]=\delta^{(n)}(t)~.
\eeq  
We note that Eq.~(22) is also a solution to Eq.~(19), but in the sense of a distribution, meaning that both the left and right hand sides of Eq.~(19) must be taken under an integral sign, so that 
$F(t) = \delta^{(n)}(t)$ acts on some arbitrary test function.

The  message to be taken from this analysis is that  the existence of the forward transform must be ensured, over and above any formal solution to the ODE.  This is a critical point,
as it serves to warn adopters of this method to be cautious in its application, since as we have just seen, the function $F(t)=C/t^{n+1}$, does not guarantee that the image, $f(s)$, exists, in spite of being
a formal solution to Equation (19).
\mbox{}\\\\
{\bf Example 3}.
Another useful, tabulated, image function is
\beq
f(s) = \exp(-n s)~,~~~~n \geq 0~.
\eeq
The ODE satisfied by $f(s)$ is given by
\beq
f^{(1)}(s) + n f(s) = 0~,
\eeq
and the ODE statisfied by $F(t)$ is therefore
\beq
-t F(t) + n F(t)=0~.
\eeq
As in Example 2 above, the non-trivial solution to this equation can only be satisfied by the Dirac-distribution, {\it viz.,}
\beq
F(t) = \mathcal{B}^{-1}[\exp(-n s)] = \delta(t-n)~.
\eeq
\mbox{}\\
{\bf Example 4}.~Our final example for this subsection is to find the two-sided ILT of
\beq 
f(s) = \frac{e^{-k/s}}{s}~,~~~~k\geq 0~.
\eeq
The ODE satisfied by $f(s)$ is
\beq
s^2 f^{(1)}(s) +s f(s) - k f(s)=0~,
\eeq
which implies that $F(t)$ satisfies
\beq
 t F^{(2)}(t) +  F^{(1)}(t)  +k F(t)=0~.
\eeq 
It is an excellent exercise to show that the only acceptable solution 
to this second-order ODE is
\beq
F(t) = J_0(2\sqrt{k t}) \Theta(t)~.
\eeq
Once again, the simple two-sided operational rules have resulted in the correct one-sided inverse Laplace transform.
\subsubsection{A few lesser known results}
\noindent {\bf Example 5}.   Consider
\beq
f(s) = \exp({a s^3})~,
\eeq
where $a$ is a real positive constant.  This image is rarely found in tables of ILTs, but it is easily dealt with using our method.  To wit, note that  elementary differentiation yields
\beq
f^{(1)}(s) -3 a s^2 f(s) =0~.
\eeq
The term by term ILT of Eq.~(32) gives
\beq
F^{(2)}(t) + \frac{1}{3a}tF(t)=0~.
\eeq
The solution to the above second-order ODE is well-known to be given by a linear combination of Airy functions,
\beq
F(t) =  C_1 {\rm Ai}\left[ \frac{-t}{(3a)^{1/3}}\right] + C_2 {\rm Bi}\left[ \frac{-t}{(3a)^{1/3}}\right]~,
\eeq
where $C_1$ and $C_2$ are constants of integration.   Since $\bigtriangleup_n=0$ for the two-sided LT, we require that 
\beq
\lim_{t\to\pm \infty} e^{-st}F(t) = 0~,
\eeq
from which we find that $C_2=0$.  The constant $C_1$ is obtained by once again ensuring that the forward transform yields Eq.~(31), with the final result being given by
\beq
F(t) = \mathcal{B}^{-1}[\exp(a s^3)] = \frac{1}{(3a)^{1/3}} {\rm Ai}\left[ \frac{-t}{(3a)^{1/3}}\right]~.
\eeq
\mbox{}\\
{\bf Example 6}.  We now wish to find the ILT for the following images

\beq
f_F(s) = \frac{\pi k_B T}{\sin(\pi k_BTs)} = b \csc bs~,
\eeq
and 
\beq
f_B(s) = -\frac{\pi k_B T}{\tan(\pi k_B T s)} = -b \cot b s~,
\eeq 
where $k_B$ and $T$ are constants, and $b \equiv \pi k_B T$.   Equations (37) and (38) have tremendous importance in the  formulation of quantum mechanics, semi-classical~\cite{brackbhaduri} 
and nuclear physics~\cite{ring} at finite  temperature, 
as will be discussed below.
We begin with $f_F(s)$ and note that $\csc(x-\pi) = -\csc x$ , which allows us to write
\begin{equation}
f_F\left(s -\frac{\pi}{b}\right) = -f_F(s).
\end{equation}
We next multiply both sides of Eq.~(39) by $s - \frac{\pi}{b}$ to obtain
\begin{equation}
\left(s - \frac{\pi}{b}\right)f_F\left(s - \frac{\pi}{b}\right) = -\left(s - \frac{\pi}{b}\right)f_F(s)~,
\end{equation}
after which a term by term ILT yields 
\begin{equation}
(e^{\frac{t \pi}{b}} +1)F_F^{(1)}(t)=\frac{\pi}{b}F_F(t)~.
\end{equation}
The solution to Eq.~(41) is formally given by 
\begin{equation}
F_F(t) = \exp \left(\frac{\pi}{b} \int\frac{dt}{e^{t \pi /b} +1)}\right).
\end{equation}

It is straightforward to show that 
\begin{equation}
F_F(t) = \frac{e^C}{e^{-t\pi/b}+1} = \frac{e^C}{e^{-t/k_BT}+1}~,
\end{equation}
where $C$ is a constant of integration.  
We may fix $C$ by appealing to final value theorem for the LT, namely,~\cite{vanderpol}
\beq
\displaystyle \lim_{t \to \infty} F(t) =  \displaystyle \lim_{s \to 0}  s f(s)~, 
\eeq
which in the present case gives
\beq
e^C= \displaystyle \lim_{s \to 0} \frac{bs}{\sin(bs)} = 1~.
\eeq
Finally, it follows that 
\begin{equation}
 F_F(t) = \mathcal{B}^{-1}\left[\frac{\pi k_B T}{\sin(\pi k_B T s)} \right]= \frac{1}{e^{-t/k_BT}+1}~.
 \end{equation}
We have only seen a similar form of Eq.~(46) explicitly tabulated in one reference, where it is simply stated without derivation.~\cite{vanderpol}  
 
We now move onto 
\beq
f_B(s) = -\frac{\pi k_B T}{\tan\pi k_B T s}~.
\eeq 
Making use of the fact that $\cot(x-\pi) = \cot x $, we readily obtain
  \begin{equation}
 f_B\left(s - \frac{\pi}{b} \right)= f_B(s)~,
\end{equation}
which after multiplication on both sides by $s -\frac{\pi}{b}$ yields
\begin{equation}
\left(s -\frac{\pi}{b}\right)f_B\left(s - \frac{\pi}{b}\right)= \left(s - \frac{\pi}{b}\right)f_B(s).
\end {equation}
The term by term ILT of Eq.~(49) gives
\begin{equation}
(e^{t\pi/b}-1)F_B^{(1)}(t)=-\frac{\pi}{b}F_B(t)~.
\end{equation} 
Analogous to the treatment of $F_F(t)$ above, the solution to this ODE is given by

\begin{equation}
F_B(t) = C\exp\left(\frac{\pi}{b}\int \frac{dt}{1 - e^{t\pi/b}}\right)= \frac{C}{e^{-t/k_B T}-1}~.
\end{equation}
Another application of the final value theorem shows that $C=1$ and therefore~\cite{vanderpol}
\begin{equation}
F_B(t) = \mathcal{B}^{-1}\left[-\frac{\pi k_B T}{ \tan(\pi k_B T s)}\right] = \frac{1}{e^{-t/k_BT}-1}~.
\end{equation}
\subsubsection{Physical Application: Ideal harmonically trapped quantum gases at finite temperature}
An immediate application of the inverse transforms derived above are harmonically trapped, ideal quantum gases at finite temperature.  
A self-contained discussion of the physics of these systems is well beyond the scope
of this article, so here, we wish to simply provide the basic mathematical tools, and introduce how the method we have proposed can be used to introduce a complementary formulation of
quantum systems, that is not widely known to many physicists.  The interested reader is encouraged to examine Refs.~[\onlinecite{brackbhaduri},\onlinecite{vanzyl,vanzyl2}] and the references therein for more details.

The central theoretical tool from which all thermodynamic properties of the ideal gas can be derived is the Bloch density matrix (BDM).  At zero temperature, the BDM is given by
\beq
C_0(\rv,\rv ';s) = \sum_{\rm all~i}\psi_i^\star(\rv)\psi_i(\rv')\exp(-s\epsilon_i)~,
\eeq
where the $\psi_i$'s and the one-particle energies $\epsilon_i$ are the solutions to the time-independent Schr\"odinger equation, and the quantity, $s$, is a complex variable.  The BDM appears to require the explicit single-particle wave functions for its evaluation, but in fact, this is not always the case.  The BDM satisfies the Bloch equation
\beq
H_r C_0(\rv,\rv';s) = -\frac{\partial C_0(\rv,\rv';s)}{\partial s}~,
\eeq
where $H_r$ is the Hamiltonian and $C_0(\rv,\rv';s)$ is subject to the initial condition
\beq
C_0(\rv,\rv';s=0) = \delta(\rv-\rv')~.
\eeq
Therefore, an explicit form for $C_0$ is, at least in principle, possible once the external potential is defined, and used in the Hamiltonian, $H_r$, to solve Equation (54).  

In the present application, we will take the external confinement to be a three-dimensional isotropic harmonic oscillator, $V(r) = \frac{1}{2}m \omega_0^2 r^2$, with $r=\sqrt{x^2+y^2+z^2}$, and
$\omega_0$ the trapping frequency.  Under such a confining potential, the {\em exact} zero temperature BDM is found to be given by~\cite{vanzyl}
\beq
C_0(\rv,\rv';s)= \left( \frac{1}{2\pi \sinh(s)}\right)^{3/2}\exp\left[
-\left(\frac{\rv + \rv'}{2}\right)^2\tanh(s/2) - \left(\frac{\rv - \rv' }{2}\right)^2\coth(s/2)\right]~.
\eeq
The zero temperature BDM is independent of the quantum statistics of the particles, meaning that we have a unified treatment for either Fermi or Bose gases (including Bose-Einstein Condensation).
Using Eq.~(56), we may obtain the {\em finite temperature} first-order density matrix (FDM) {\it via} a two-sided ILT,~\cite{brackbhaduri,vanzyl,vanzyl2} 
\beq
\rho(\rv,\rv';T) = \mathcal{B}_{\mu}^{-1}\left[ \frac{g}{s}C_T(\rv,\rv';s)\right]~,
\eeq
where
\begin{eqnarray}
C_T(\rv,\rv';s) &=& C_0(\rv,\rv';s)\frac{\pi s k_B T}{\sin(\pi s k_B T)}~~~~~~~({\rm fermions})~,\nonumber \\
&=& -C_0(\rv,\rv';s)\frac{\pi s k_B T}{\tan(\pi s k_B T)}~~~~({\rm bosons})~,
\end{eqnarray}
and now $k_B$ and $T$ take on the physical significance of the Boltzmann constant and temperature, respectively.  Note that in Eq.~(57), we have introduced the subscript, $\mu$, in the two-sided ILT to emphasize
that the conjugate variable to $s$ in this problem is $\mu$, the chemical potential.  The factor, $g$ in Eq.~(57) depends on the spins of the particles through $g=2s_1+1$, where $s_1$ is the spin of the particle.  For spin-$1/2$ fermions, $g=2$, and for spinless bosons, $g=1$.
The finite temperature BDM, $C_T$, encodes the quantum statistics by appropriately weighting the
zero temperature Bloch density matrix.
It should now be evident why we must utilize the two-sided ILT, rather than the one-sided ILT; the chemical potential, $\mu$, must
be allowed to take on both positive and negative values at finite temperature, whereas the one-sided ILT only allows for $\mu \geq 0$.

Equation (57) is the fundamental quantity from which all of the thermodynamic properties of the trapped gases (fermions or bosons) may be investigated.  For example,
the finite temperature spatial density, $\rho(\rv;T)$ is immediately obtained by putting $\rv=\rv'$ in Eq.~(57), while the 
the finite temperature kinetic energy density can be found from
\beq
\tau(\rv;T) = \sum_{i=1}^{3} \left[
\frac{\partial^2}{\partial x_i \partial x_i'} \rho(\rv,\rv';T)\right]~.
\eeq
Following the calculations in Ref.~[\onlinecite{vanzyl2}], the ILTs we have presented earlier may be used to explicitly evaluate $\rho(\rv,\rv';T)$.   

In the case where the potential, $V(\rv)$, is such that an exact, closed form expression for $C_0(\rv,\rv ';s)$  is not possible, an $\hbar$-expansion may be performed, {\it viz.,} 
\begin{equation}
 C_0(\rv,\rv' ;s) = \left (\frac{1}{\lambda}\right)^3 e^{-s V(\rv) }e^{-\frac{m (\rv - \rv')^2}{2\hbar^2 s}}
 \left(
 1 - \frac{\hbar^2 s^2}{12 m}\left[ \nabla^2 V - \frac{s}{2}(\nabla V)^2\right] + \cdot \cdot \cdot
 \right)~,
 \end{equation}
where for economy of space, we have only shown terms up to relative order $\hbar^2$, $\lambda \equiv (2\pi\hbar^2 s/m)^{1/2}$ and $\hbar$ is Planck's constant divided by $2\pi$.  
Equation (60) is called the semi-classical expansion of the BDM,
and the higher order $\hbar$-terms, containing gradients of the potential, represent quantum corrections to the classical result. 
The semi-classical expansion may be used as the foundation for the study of quantum effects in a variety of systems by systematically introducing higher-order quantum corrections and examing their
influence on the physical properties of the system.  A detailed discussion of the applications of Eq.~(60) in semi-classical physics may be found in Reference [\onlinecite{brackbhaduri}].
 
\subsection{Inverse Fourier transform}
The inverse Fourier transform (IFT) is mathematically much easier than the ILT, since the IFT is an integral over the real line.  Nevertheless, it is worthwhile illustrating our method as it applies to the IFT, particularly in 
cases where  standard approaches to finding the inverse transform require some ``tricks''  for its evaluation.  
\mbox{}\\\\
{\bf Example 7}.
Let us then consider
the following image
\begin{equation}
f(s)=\sqrt{\frac{\pi}{\sigma}}e^{\frac{-s^2}{4\sigma}}~.
\end{equation}
Simple differentiation shows that
\begin{equation}
f^{(1)}(s)=\left(\sqrt{\frac{\pi}{\sigma}}e^{\frac{-s^2}{4\sigma}}\right)\left(-\frac{s}{2\sigma}\right)
                     =-\frac{s}{2\sigma}f(s)~,
\end{equation}
and thus $f(s)$ satisfies the following first order differential equation
\begin{equation}\label{ode}
f^{(1)}(s)+\frac{s}{2\sigma} f(s)=0~.
\end{equation}
Applying the properties in Table II to Eq.~(63) above leads to the following ODE for $F(t)$ 
\begin{equation}
F^{(1)}(t) + {2\sigma}tF(t)=0~.
\end{equation}
Obtaining the solution to Eq.~(64) is elementary, and is given by
\begin{equation}
F(t) = C e^{-\sigma t^2}~.
\end{equation}
The constant $C$ is adjusted such that the Fourier transform of $F(t)$ is $f(s)$. It then follows that
\begin{equation}
F(t) = \mathcal{F}^{-1}\left[\sqrt{\frac{\pi}{\sigma}}e^{\frac{-s^2}{4\sigma}}\right]=e^{-\sigma t^2}~.
\end{equation}
Obtaining Eq.~(66) using standard techniques is more involved.~\cite{griffiths}

\subsection{Mellin transform}
The Mellin transform (MT) may also be treated by the ODE method presented here.  The Mellin transform is defined by~\cite{tit}
\beq
f(s) = \mathcal{M}[F(t)] = \int_{0}^{\infty} t^{s-1} F(t)~dt~,
\eeq
and its inverse is given by
\beq
F(t) = \mathcal{M}^{-1}[f(s)] = \frac{1}{2\pi i}\int_{c -i\infty}^{c+i\infty}t^{-s} f(s)~ds~.
\eeq
Once again, we observe that the two-sided LT may be viewed as the fundamental tool for our operational calculus, since the MT may also be obtained 
from to the two-sided LT by noting that~\cite{vanderpol}
\beq
\mathcal{M}[F(t)] = \mathcal{B}[F(e^{-t})]~.
\eeq
The only thing left to do is to determine the corresponding ``rules'' as we did for
transforms with exponential kernels.  It is straightforward to confirm that following rules are obeyed for the Mellin transform:
\begin{eqnarray}
\mathcal{M}[F(at)] &=& a^{-s} f(s)\nonumber \\
\mathcal{M}[t^a F(t)] &=& f(s+a)\nonumber \\
\mathcal{M}[F^{(1)}(t)] &=& -(s-1) f(s-1)\nonumber \\
\mathcal{M}[t^n F^{(n)}(t)] &=& (-1)^n \frac{\Gamma(n+s)}{\Gamma(s)} f(s)~.
\end{eqnarray}
\mbox{}\\
{\bf Example 8}.
To illustrate the method applied to the MT, let us use it to evaluate the complex contour integral,
\beq
 \frac{1}{2\pi i}\int_{c -i\infty}^{c+i\infty}t^{-s} \Gamma(s)~ds~,
 \eeq
 which we recognize as the inverse Mellin transform of $f(s)=\Gamma(s)$.  According to the rules given above, we note that
 \beq
 f(s+1) =s f(s)~.
 \eeq
 Applying the inverse rules to Eq.~(72) yields
 \beq
 F^{(1)}(t) = -F(t)~,
\eeq
which is immediately seen to to have the solution
\beq
F(t) = Ce^{-t}~.
\eeq
The constant, $C$, is found by ensuring that the forward transform, $f(s)=\Gamma(s)$ (which is just a {\em real} integral), is obtained, which gives $C=1$.
Therefore, we have evaluated the complex contour integral,
\beq
e^{-t} =  \frac{1}{2\pi i}\int_{c -i\infty}^{c+i\infty}t^{-s} \Gamma(s)~ds~~,
 \eeq
using only elementary methods.
 
\section{Multipole expansions of functions}

In this section, we wish to illustrate how to obtain the ``multipole'' expansion of a function, $F(t)$, in the form of a series involving the Dirac-delta distribution and its derivatives.  Such expansions 
have been found to be useful in semi-classical physics, where they may be used to introduce quantum corrections,~\cite{airy,heller,hupper} and in nuclear physics, for the calculation of approximate wave functions
for realistic nuclear potentials.~\cite{lukyanov}

The method presented here is general, and uses a very simple approach based on the two-sided inverse Laplace transform.  An immediate benefit  of our formulation is that it
permits an easy determination of {\em all} of the moments of a function, $F(t)$ (see Eq.~(79) below).  In addition, our approach allows us to obtain results that have
previously required intense mathematical analysis  and dedicated papers for their derivation. 

In the most generic of cases, we begin with the forward two-sided LT of $F(t)$, {\it viz.,}
\begin{equation}
f(s) = \int_{-\infty}^{\infty} e^{-st}F(t)dt~.
\end{equation}
Assuming that $f(s)$ is analytic, we use the series expansion,
$e^{-st}=\sum_{n=0}^{\infty} (-1)^ns^nt^n/n!$, in Eq.~(76) to obtain
\begin{equation}
f(s) = \sum_{n=0}^{\infty}\left(\frac{(-1)^n}{n!}\int_{-\infty}^{\infty} t^nF(t)dt\right) s^n.
\end{equation}
At this point, we are not worrying about the fact that we have taken the summation outside of the integral, since our end result will be an expansion that is defined only in the
sense of a distribution, and therefore will only be valid under the integral sign.
We now term by term Laplace invert Eq.~(77), which yields
\begin{equation}\label{asymp}
F(t) = \sum_{n=0}^{\infty}\frac{(-1)^n}{n!}I_n \delta^{(n)}(t)~,
\end{equation}
where
\beq
I_n=\int_{-\infty}^{\infty} q^nF(q)~dq~,
\eeq
is the $n$-th moment of the function, $F(t)$.  The $n=0$ term is sometimes referred to as the  the ``monopole'' moment, and the $n=1$ the ``dipole'' moment, and so on.
Equation (78) is the general form of the multipole moment expansion, and is to be viewed in the distribution sense, having meaning only under the integral sign. 
\mbox{}\\\\
{\bf Example 9}.~In most practical applications, the moments of the function, Eq.~(79), need not be explicitly calculated.  To see this, let us first consider $F(t) = {\rm Ai}(t)$.  
We can deduce from Eq.~(36) that,
\beq
f(s) = \exp\left(\frac{-s^3}{3}\right)~,
\eeq
is the two-sided LT of $F(t)$.  We then formaly write $f(s)$ as a series expansion,
\beq
\exp\left(\frac{-s^3}{3}\right) =  \sum_{n=0}^{\infty} \frac{(-1)^ns^{3n}}{3^nn!}~,
\eeq
and then term by term ILT both sides of Eq.~(81) (with the aid of Example 2) to obtain
\beq
{\rm Ai}(t) =\mathcal{B}^{-1}\left[ \sum_{n=0}^{\infty} \frac{(-1)^ns^{3n}}{3^nn!}\right]=\sum_{n=0}^{\infty} \frac{(-1)^n\delta^{(3n)}(t)}{3^nn!}~.
\eeq
A comparison of Eq.~(79) with Eq.~(82) may be used to determine the moments of the Airy function.  Equation (82) has  been derived earlier by several authors,~\cite{airy,heller,hupper} but by using 
a much more involved approach than the one presented here. As commented above, it is important to remember
that Eq.~(82) is to be interpreted strictly in the sense of a distribution.
\mbox{}\\\\
{\bf Example 10}.~Next we find an analogous expansion for the Bessel function, $F(t) = J_{\nu}(\omega t)\Theta(t)$, where ${\rm Re}~[\nu] > -1$. 
The well known Laplace transform of this function is~\cite{grad}
\begin{equation}
f(s) = \mathcal{B}[ J_{\nu}(\omega t)\Theta(t)] = \frac{\left(\sqrt{\omega^2+s^2}-s\right)^{\nu}}{\omega^{\nu}\sqrt{\omega^2+s^2}}.
\end{equation} 
As in the previous example, we consider a formal series expansion of $f(s)$,
\begin{equation}
 \frac{\left(\sqrt{\omega^2+s^2}-s\right)^{\nu}}{\omega^{\nu}\sqrt{\omega^2+s^2}} = \sum_{n=0}^{\infty} \frac{2^n \Gamma\left(\frac{n+1-\nu}{2}\right)s^n}{n!\Gamma\left(\frac{-\nu-n+1}{2}\right)\omega^{n+1}}~,
\end{equation}
followed by a term by term two-sided ILT of both sides of Equation (84).  The result of this inversion is 
\begin{equation}
J_{\nu}(\omega t)\Theta(t)= \sum_{n=0}^{\infty} \frac{2^n \Gamma\left(\frac{n+1-\nu}{2}\right)\delta^{(n)}(t)}{n!\Gamma\left(\frac{-\nu-n+1}{2}\right)\omega^{n+1}}~.
\end{equation}
To our knowledge, Eq.~(85) is a new result, and its use can immediately be seen from an examination of integrals for the form
\begin{eqnarray}
I(\nu) &=& \int_{0}^{\infty} \varphi(r) J_{\nu}(r)~dr\nonumber \\
&=& \sum_{n=0}^{\infty} (-1)^n\frac{2^{n-1} \Gamma\left(\frac{n+1-\nu}{2}\right)\varphi^{(n)}(0)}{n!\Gamma\left(\frac{-\nu-n+1}{2}\right)}~.
\end{eqnarray}
Again, a comparison of Eq.~(79) with Eq.~(85) allows us to obtain {\em all} of the moments of $J_{\nu}$, {\it viz.,}
\beq
I_n = (-1)^n \frac{2^n \Gamma\left(\frac{n+1-\nu}{2}\right)}{\Gamma\left(\frac{-\nu-n+1}{2}\right)\omega^{n+1}}~,
\eeq
which by direct application of Eq.~(79) would have been more difficult.
\mbox{}\\\\
{\bf Example 11}.~As our final example, we consider the multipole expansion of the Fermi function,~\cite{lukyanov}
\beq
F(t)=\frac{1}{[e^{(t-R)/a}+1]}~.
\eeq
In nuclear physics applications, $R$ would denote the nuclear radius, and $a$ the ``diffuseness'' of its surface.  It is straightforward to deduce from Eq.~(46), and property  IV of Table II, that $F(t)$ has the image
\beq
f(s) = -\frac{\pi a}{\sin(\pi a s)} e^{-R s}~.
\eeq
We next expand $f(s)$ as a power series about the diffuseness parameter, $a$, and obtain the expansion
\beq
-\frac{\pi a}{\sin(\pi a s)} e^{-Rs} = -\frac{e^{-R s}}{s} - \sum_{k=0}^{\infty} A_{2k+1} a^{2k+2} e^{-R s}s^{2k+1}~,
\eeq
where
\beq
A_{2k+1} = \frac{2(2^{2k+1}-1)\pi^{2k+2}}{(2k+2)!}|B_{2k+2}|~,
\eeq
and the $B_{2k+2}$ are the known Bernoulli numbers.  The term by term two-sided ILT of Eq.~(90) gives
\beq
\frac{1}{[e^{(t-R)/a}+1]} = \Theta(R-t) - \sum_{k=0}^{\infty} A_{2k+1} a ^{2k+2} \delta^{(2k+1)}(t-R)~.
\eeq
Equation (92), which we have derived in only a few lines, has previously been the focus of an {\em entire} paper by L.~Lukyanov,~\cite{lukyanov} thereby emphasizing just how useful the approach we present here can be in a wide array of problems.
In fact, what is remarkable is just how trivial Eq.~(92) was to obtain when compared to the analysis presented in Reference [\onlinecite{lukyanov}].
\section{Closing Remarks}
We have presented an alternative approach for the evaluation of common inverse transforms occurring in physics without having to introduce complex contour integration.  
Our operational calculus has been
based on the two-sided Laplace transform, which has allowed us to examine a much wider class of functions than would be possible using other transforms.   
We have also highlighted the
power of the two-sided inverse Laplace transform by demonstrating how it can often make trivial work of problems ({\it e.g.,} Examples 9 and 11), which otherwise require a much more extensive analysis.  

Some may find the presentation in this tutorial lacking in mathematical rigor.  In particular, we have purposefully avoided technical discussions of strips of convergence, for which at best, the integral transform is valid. 
For the examples we have provided, where the strip of convergence is not indicated, the latter has to be investigated, given the original $F(t)$, for each individual case separately.  Indeed,  a useful exercise to 
give to students would be to determine
the strips of convergence for each of the examples provided in this paper.

Nevertheless, it is important to keep in mind that in undergraduate physics culture,  advanced mathematical tools ({\it e.g.,} integral transforms)
are often needed right away --- students just need to know {\em how to use} the tools,  with the details of all the theorems and proofs being covered in more formal mathematics courses.   Therefore, we  hope 
the techniques presented in this tutorial will also find their way into the physics classroom, where they may be used to enable students to evaluate  integral transforms with confidence, and without being slaves to tables.  
We also hope that
our presentation will inspire students to further examine the formulations of quantum mechanics, semi-classical physics,~\cite{brackbhaduri} and density-functional theory,~\cite{brackbhaduri,farrell} all of which may be based on the two-sided Laplace transform.
Finally, we believe that our general results for the multipole expansion of functions will be of interest to practicing physicists, particularly those involved in the foundations of quantum, semi-classical and nuclear physics research.
\section{acknowledgements}
Z. MacDonald would like to acknowledge the Natural Sciences and Engineering Research Council of Canada
(NSERC) USRA program for financial support.  B. P.  van Zyl and A. Farrell would also like to acknowledge the NSERC Discovery Grant program for additional
financial support.

\end{document}